\documentclass[preprint,preprintnumbers,amsmath,amssymbm,a4paper]{revtex4}
\usepackage{graphicx}
\usepackage{dcolumn}
\usepackage{bm}
\usepackage{color}%
\usepackage{pdfpages}%
\begin{document}

\title{Re-examination of electronic structure of dilute Kondo transition-metal ions substituted into a Heavy Fermion compound}

\author{Kou Takubo}
\email{ktakubo@aoni.waseda.jp}
\affiliation{Department of Applied Physics, Waseda University, Tokyo 169-8555, Japan}
\affiliation{Institute for Solid State Physics (ISSP), University of Tokyo, Kashiwa, Chiba 277-8581, Japan}
\author{Shintaro Suzuki}
\affiliation{Institute for Solid State Physics (ISSP), University of Tokyo, Kashiwa, Chiba 277-8581, Japan}
\author{Kohei Yamamoto}
\affiliation{Institute for Solid State Physics (ISSP), University of Tokyo, Kashiwa, Chiba 277-8581, Japan}
\affiliation{NanoTerasu Center, National Institutes for Quantum Science and Technology, Sendai, Miyagi 980-8572, Japan}
\author{Kohei Yamagami}
\affiliation{Japan Synchrotron Radiation Research Institute, Sayo, Hyogo 679-5198, Japan}
\author{Masafumi Horio}
\affiliation{Institute for Solid State Physics (ISSP), University of Tokyo, Kashiwa, Chiba 277-8581, Japan}
\author{Toshiaki Ina}
\affiliation{Japan Synchrotron Radiation Research Institute, Sayo, Hyogo 679-5198, Japan}
\author{Kiyohumi Nitta}
\affiliation{Japan Synchrotron Radiation Research Institute, Sayo, Hyogo 679-5198, Japan}
\author{Masaichiro Mizumaki}
\affiliation{Faculty of Science, Kumamoto University, Kurokami Chuou-ku, Kumamoto 860-8555, Japan}
\author{\\Eiji Ikenaga}
\affiliation{Japan Synchrotron Radiation Research Institute, Sayo, Hyogo 679-5198, Japan}
\author{Yosuke Matsumoto}
\affiliation{Institute for Solid State Physics (ISSP), University of Tokyo, Kashiwa, Chiba 277-8581, Japan}
\author{Hiroki Wadati}
\affiliation{Institute for Solid State Physics (ISSP), University of Tokyo, Kashiwa, Chiba 277-8581, Japan}
\affiliation{
Department of Material Science, Graduate School of Science, University of Hyogo, Ako, Hyogo 678-1297, Japan}
\affiliation{Institute of Laser Engineering, Osaka University, Suita, Osaka 565-0871, Japan}
\author{Satoru Nakatsuji}
\affiliation{Department of Physics, University of Tokyo, Bunkyo-ku, Tokyo 113-0033, Japan}
\affiliation{Institute for Solid State Physics (ISSP), University of Tokyo, Kashiwa, Chiba 277-8581, Japan}
\affiliation{Trans-scale Quantum Science Institute, University of Tokyo, Bunkyo-ku, Tokyo 113-0033, Japan}
\affiliation{Institute for Quantum Matter and Department of Physics and Astronomy, Johns Hopkins University, Baltimore, MD 21218, USA}
\date{\today}

\begin{abstract}

Correlations between the localized and conductive spins/charges have been the central issue of various fascinating quantum phenomena found on itinerant electron systems. Here, the obvious multiplet structures are presented on the Mn 2$p$ to 3$d$ x-ray absorption for a heavy fermion $\alpha$-(Yb,Lu)(Al$_{1-x}$Mn$_x$)B$_4$, indicating that the unoccupied electronic structure of the Mn site is described as the correlated high-spin 2+, even though magnetic measurements show the Mn sites to be nonmagnetic. This apparently paradoxical result demonstrates that a ligand field can effectively appear between localized Mn 3$d$ and surrounding B 2$p$ orbitals, which has been anticipated as a manifestation of a Kondo effect but not been clearly confirmed for most itinerant metals in spectroscopy. By contrast, the Mn 2$p$ photoemission indicates that the occupied Mn$^{2+}$ 3$d$ electrons still exhibit itinerant and nonlocally screened nature also owing to the Kondo-like correlation with the conductive B 2$p$, and heavier Yb 4$f$ and 5$d$ bands below the Fermi energy. The asymmetry on the particle-hole stimulates a reconsideration of the correlation and screening effects in the core-level spectroscopies.

\end{abstract}

\maketitle
Heavy fermion (HF) has been a long-standing problem in correlated electron physics, offering key insights into the emergent quantum nontrivial phenomena such as quantum criticality, unconventional superconductivity, and exotic magnetic/electronic fluctuations \cite{YoshidaTheory}.
Interaction and hybridization between conduction electrons and localized moments on the partially-filled orbitals of transition-metal (TM) $d$, rare-earth 4$f$, and actinide 5$f$ have widely been argued as the origin of various HF behaviors.
The HF behavior and Kondo effect were first proposed for diluted 3$d$ TM alloys \cite{YoshidaTheory, Kondo64,Yoshida65,Anderson61}. Also, many recent discovered 4$f$- and 5$f$-based HFs such as
CeCoIn$_5$ , URu$_2$Si$_2$ \cite{URu2Si2_1,URu2Si2_2}, Ce$Tr_2$Al$_{10}$ ($Tr$ = Fe, Ru, and Os) \cite{CeTr2Al10_1,Ishiga14}, and $\alpha$-YbAl$_{1-x}$(Fe,Mn)$_{x}$B$_4$ (Fig. 1) \cite{Suzuki18,Kuga18} contain TM sites that strongly influence their HF and magnetic properties. In this regard, detailed, site-selective studies of the electronic structure at TM sites are highly desirable for understanding both their magnetic and electric properties in HF materials as well as those of the rare-earth or actinide sublattices.

\begin{figure}
	\includegraphics[width=1\linewidth]{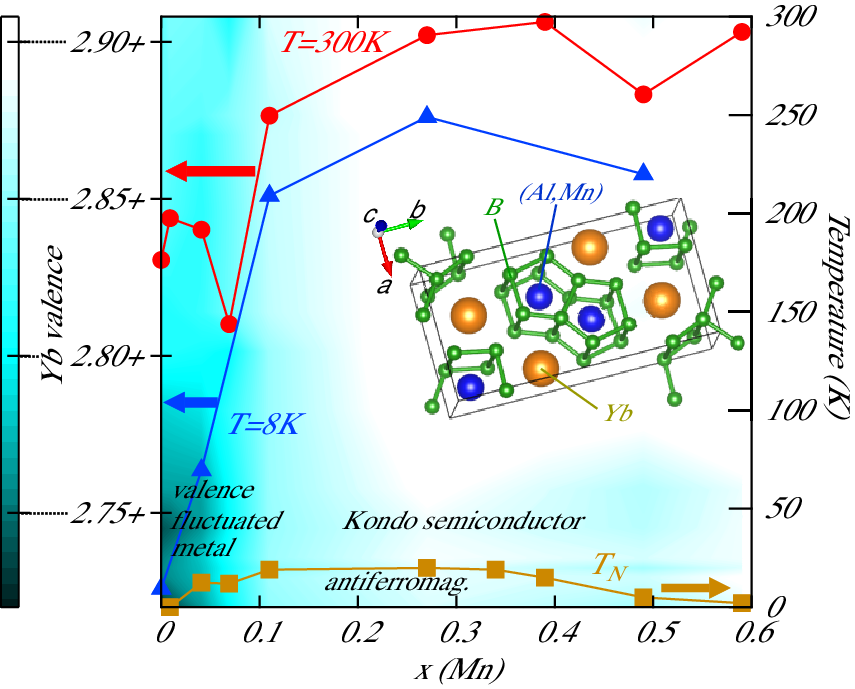}%
	\caption{
		Relationship between the phase diagram and valence of the Yb sites for $\alpha$-YbAl$_{1-x}$Mn$_x$B$_4$.
		Blue triangles and red circles indicate valence of the Yb sites estimated from XAS at Yb $L_3$ edge at $T$ = 8 and 300 K, respectively (left axis). 
		Shaded background of the figure indicates the valence (scale in left) at the corresponding temperature (right axis).
		More details of temperature and composition dependence are given in SM \cite{supp}.
		Brown squares indicate $T_\mathrm{N}$ reported in the previous study (right axis) \cite{Suzuki18}.
		Inset shows the crystal structure.
	}
\end{figure}

Core‑level spectroscopies, such as x‑ray absorption spectroscopy (XAS) and x‑ray photoemission spectroscopy (XPS), are most powerful probes of the element‑specific electronic structure in correlated materials.
In relatively localized electron systems — for example, 3$d$ TM oxides \cite{Groot90,Bocquet92,Mitra03,Biesinger11} and borides \cite{MnB24} including high‑$T_{\rm{C}}$ cuprates,
\cite{Taguchi05}, colossal magnetoresistive manganites \cite{Horiba04,Takubo07}, and a few TM-oxide-based HFs \cite{Ca3Cu3RuO12_1,Ca3Cu3RuO12_2,Ca3Cu3RuO12_3} — core-level 2$p$ XAS and XPS commonly exhibit pronounced multiplet and satellite features.
These structures in XAS and XPS reflect the ligand crystal-field (CF) and configuration‑interaction (CI) in the final states of the TM $d$ electrons for unoccupied and occupied states, respectively, and are routinely analyzed in detail using cluster‑model calculations based on the Anderson impurity model or ligand‑field theory  \cite{Bocquet92,Groot90,Taguchi05,Ca3Cu3RuO12_1,Ca3Cu3RuO12_2,Ca3Cu3RuO12_3,Haverkort12,Crispy,Hufner}.
By contrast, in more itinerant systems — including many of HFs  \cite{Steiner80, Kimura11,CeCoSn}, and iron‑based superconductors \cite{Bondino08,Yang09,Kurmaev09,Oiwake13,Takubo17_1} — core‑level spectra of doped or substituted metal ions often resemble those of simple metals: broad peaks with weak fine structure and an extended high‑energy tail, a behavior that can be interpreted within an extended Fermi‑liquid framework \cite{CeCoSn,Hufner}.
Consequently, valence estimates in such systems have often been made phenomenologically from peak or edge shifts.
In practice, however, a clear spectroscopic criterion separating localized and itinerant behavior has not been established.
Moreover, the band structures of many HF systems near Fermi-energy $E_\mathrm{F}$ themselves are well described by Anderson‑type Hamiltonians \cite{YoshidaTheory,Anderson61}.

Multiplet features at core‑level edges for dilute TM ions have rarely been observed in HFs by XAS or XPS.
A likely reason is that both unoccupied and occupied $d$ states in the final states often lie close to $E_\mathrm{F}$ and can exhibit approximate particle–hole symmetry, leading to strong coupling with the highly conductive bands characteristic of HFs.
For instance, in the prototypical dilute Kondo alloy Al–Mn, Mn is theoretically expected to adopt a 2+ valence that hybridizes with conduction electrons, yet the system remains nonmagnetic due to the Kondo effect \cite{Anderson61,YoshidaTheory}.
Experimentally, Mn 2$p$ XPS in Al–Mn shows only slight deviations from metallic spectra \cite{Steiner80} and differs markedly from spectra of correlated Mn oxides \cite{Mitra03,Biesinger11}, whereas Mn $K$‑edge XAS indicates a positive Mn valence \cite{Sadoc86,Dashpande90}. Such apparent contradictions in core‑level spectroscopy are sometimes encountered in itinerant but strongly correlated systems \cite{Song08,CeCoSn}.

Among HFs, the 4$f$-electron-based $\beta$-YbAlB$_4$ is notable as the first HF superconductor that exhibits unconventional quantum criticality without external tuning \cite{Nakatsuji08,Matsumoto11,Tomita15,Okawa10}.
By contrast, its isostructural polymorph, $\alpha$-YbAlB$_4$, has a Fermi‑liquid ground state while showing similar Kondo‑lattice behavior at higher temperatures \cite{Matsumoto11_2}.
Recent work has shown that small substitutions of TM such as Fe or Mn for the Al sites can induce quantum‑critical behavior in $\alpha$-type,
similar to that of $\beta$-type \cite{Kuga18,Suzuki18}.
Another striking feature of TM‑substitution is the unusually high N\'eel temperature ($T_\mathrm{N}$): for $\alpha$-YbAl$_{1-x}$Mn$_x$B$_4$, antiferromagnetism with Kondo‑semiconductor–like behavior appears for $x>$ 0.07, and $T_\mathrm{N}$ reaches $\sim$20 K at $x$ = 0.27 (Fig. 1) \cite{Suzuki18}.
This $T_\mathrm{N}$ is the highest among Yb‑based HFs at ambient pressure, despite the fact that the Mn sites remain nonmagnetic on the muon‑spin timescale \cite{Suzuki18}.
Therefore, site‑selective electronic‑structural studies of Mn and Yb in $\alpha$-YbAl$_{1-x}$Mn$_{x}$B$_4$ are essential to elucidate the mechanisms behind the high transition temperature and the unconventional HF behavior.

In this study, we investigate the electronic structure of Mn and Yb sites in $\alpha$-(Yb,Lu)Al$_{1-x}$Mn$_{x}$B$_4$ using site‑selective x‑ray absorption spectroscopy (XAS) and hard x‑ray photoemission spectroscopy (HAXPES).
XAS and HAXPES at the Mn 2$p$ edge yield apparently contrasting results; nevertheless, both techniques reveal distinct aspects of a Kondo‑like correlation at nominally nonmagnetic, yet strongly correlated, Mn$^{2+}$ sites. We show that these observations are closely linked to a breaking of particle–hole symmetry in the correlated electronic structure on the Yb-based system.

Single crystals of $\alpha$-(Yb,Lu)Al$_{1-x}$Mn$_{x}$B$_4$ were grown by using the Al flux growth technique \cite{Suzuki18}.
XAS measurements at the Yb $L$ (2$p\rightarrow$ 5$d$) and Mn $K$ (1$s\rightarrow$ 3$d$, 4$s$, and 4$p$) absorption edges were performed at the BL01B1 beamline of SPring-8.
The absorption was estimated from the transmission.
XAS measurements at the Mn $L$ (2$p\rightarrow$ 3$d$) absorption edge were conducted both at the BL07LSU in SPring-8 and BL-16A in the Photon Factory, Japan.
The spectra were recorded using the total electron yield (TEY) mode, although the spectra for $x$ = 0.27 were recorded at both the TEY and fluorescence yield (FY) modes.
All the samples were cleaved just before the measurements.
In addition, the temperature dependence of XAS between $T$ = 40 and 300 K were examined but no noticeable change was observed.
HAXPES measurements were performed at the BL47XU beamline of SPring-8.
The photon energy was 7940 eV at $T$= 300 K.
The binding energy was calibrated by the Fermi edge of gold and the energy resolution was approximately 200 meV (FWHM).
The samples were cleaved \textit{in situ} to avoid surface contamination.

First, the valences of the Yb sites were evaluated by means of the Yb $L_{3}$-edge XAS \cite{Cornelius97,Matsuda07,Matsuda13},
and the results were summarized in Fig. 1.
The details of the temperature dependence with some raw spectral data are presented in the Supplementary Material (SM) \cite{supp}. 
The system transitions from the valence-fluctuated-metal to Kondo semiconductor having $T_\mathrm{N}$ across quantum critical point (QCP) at $x_\mathrm{c}$ = 0.009 \cite{Suzuki18}.
Correspondingly, a sharp valence crossover by $\sim$0.04 is observed between $x$ = 0.0 and 0.04 at $T$ = 8 K with respect to first-order valence transition across QCP, which is quantitatively similar to that observed in $\alpha$-YbAl$_{1-x}$Fe$_{x}$B$_4$ at $x_\mathrm{c}$=0.014 \cite{Kuga18}.
In addition, the valence exhibits a further increase by $\sim$0.11 for $x\sim$0.07 (between 0.04 and 0.11) at $T$ = 8 K. 
The valences for $x\geq$ 0.11 reach $\sim$2.9+ above $T$ = 20 K, close to 3+.
The total increase by $\sim$0.15 across the two-stage crossovers ($x$ = $\sim$0.009 and 0.07) for $\alpha$-YbAl$_{1-x}$Mn$_{x}$B$_4$ is much larger than that by $\sim$0.03 at the transition for $\alpha$-YbAl$_{1-x}$Fe$_{x}$B$_4$ \cite{Kuga18}.
Therefore, a drastic quantity instead of the chemical pressure or cell volume proposed for $\alpha$-YbAl$_{1-x}$Fe$_{x}$B$_4$ \cite{Kuga18} is expected for the Mn substitution in $\alpha$-YbAl$_{1-x}$Mn$_{x}$B$_4$, just relating to the origin of the highest $T_\mathrm{N}\sim$ 20 K for $x\geq$ 0.11.

To clarify this origin,
the electronic structure of the Mn sites was surveyed through XAS and HAXPES at Mn core-levels.
Figure 2 shows results of XAS at the Mn $L_{2,3}$ (2$p\rightarrow$ 3$d$) edges.
Distinct, sharp multiplet features are observed for all compositions, including macroscopically nonmagnetic $\alpha$-LuAl$_{1-x}$Mn$_x$B$_4$ ($x$=0.44) \cite{Suzuki18}, indicating strong CI and CF effects in the unoccupied Mn 3$d$ final states.
The multiplet structures are consistent between total‑electron‑yield (TEY) and bulk‑sensitive fluorescence‑yield (FY) measurements.
The spectral line shapes closely resemble those of MnO and are well reproduced by charge‑transfer multiplet calculations performed with Quanty \cite{Haverkort12} as the Mn$^{2+}$ high-spin ground state, as shown at the bottom of Fig. 2(a).
The detail of the calculation is described in SM \cite{supp}.

\begin{figure}
	\includegraphics[width=1\linewidth]{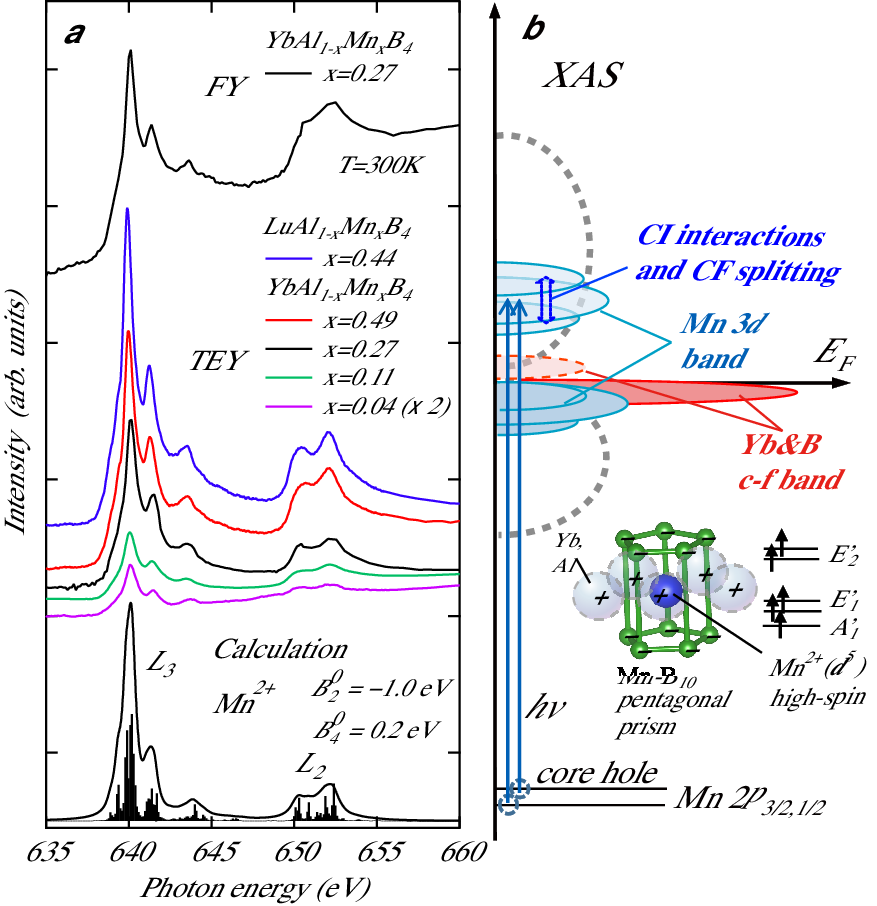}%
	\caption{XAS at the Mn $L$-edge of $\alpha$-(Yb,Lu)Al$_{1-x}$Mn$_{x}$B$_4$. (a) Spectra taken at the FY (top) and TEY (middle) modes. The bottom spectrum is obtained through a charge transfer multiplet calculation for Mn$^{2+}$ high-spin ground state \cite{supp}.
		(b) Schematic of XAS processes showing the excitation from the Mn 2$p$ core levels.
		The light blue and red parabolas indicate the density of states for the Mn$^{2+}$ 3$d$ and Yb$^{2+}$\&B $c$-$f$ bands, respectively.
		The blue bidirectional arrows indicate the CF splitting and CI of Mn 3$d$ orbitals in the unoccupied states.
		Inset indicates the Mn 3$d$ CF splitting expected in the Mn-B$_{10}$ pentagonal prism. The four and one sides of the prism face Yb and Al cations, respectively.
	}
\end{figure}

In this calculation, the best fit values of Slater integrals are reduced to 85$\%$ from the atomic Hartree-Fock values \cite{Groot90}. The Wybourne coefficients of $B_{2}^{0}$ and $B_{4}^{0}$ for spherical harmonics are $-$1.0 and 0.2 eV, respectively \cite{supp}, which describe the CF splitting in the Mn-B$_{10}$ pentagonal prism having $D_{5h}$ symmetry [Inset of Fig. 2(b)].
The slight reduced Slater integrals and considerable values of $B_{2}^{0}$ and $B_{4}^{0}$ indicate the splitting between the off-plane bonding and in-plane antibonding types of the 3$d$ orbitals, guaranteeing the hybridization or correlation between the Mn 3$d$, and surrounding B 2$p$, and/or Yb 4$f$ and 5$d$ electrons \cite{supp}.
The Mn and huge Yb cations are lies in similar $ab$ plane only about 2.9 \r{A} apart. Thus the electrons in these sites are also possible to correlate directly with each other as well as those in the B sites in the upper and lower layers \cite{supp}.

Furthermore, the valence of the Mn sites could be confirmed to be 2+ or higher but metallic phenomenologically, by the bulk-sensitive XAS at Mn $K$-edge absorption \cite{Ghasemi16,Furrer18}, 
as also given in SM \cite{supp}. 
The correlation effect at the Mn sites verified from these XAS results will reflect the strong covalency in the borides, as like those often observed in the 3$d$-TM oxides.
One may feel that these results will contradict the fact that the Mn site is nonmagnetic. However, it will be precisely consistent with the picture predicted by early theoretical researches for diluted or substituted Kondo TM ions from the Al sites in highly itinerant compounds, as discussed in the introduction \cite{YoshidaTheory,Anderson61}.
The discrepancy of the observations will arise from the difference of timescales between XAS and magnetic measurements as discussed in the XPS study for Al-Mn system \cite{Steiner80}.

Strikingly, however, high‑resolutional Mn 2$p$ HAXPES spectra of $\alpha$-(Yb,Lu)Al$_{1-x}$Mn$_x$B$_4$ (Fig. 3) show little evidence of multiplet nor charge‑transfer satellite features, despite TM 2$p$ HAXPES often being considered complementary to TM $L$-edge XAS.
The Mn 2$p$ spectra differ markedly from those of correlated Mn oxides \cite{Biesinger11,Song08}, suggesting an itinerant character of the Mn 3$d$ electrons in the HAXPES final state.
Nevertheless, the Mn 2$p_{3/2}$ and 2$p_{1/2}$ binding energies ($\sim$639.1 eV and $\sim$650.0 eV, respectively) are shifted to higher energy relative to pure Mn metal ($\sim$638.7 and $\sim$649.8 eV, respectively) \cite{pureMnHAXPES2}, and the high‑energy asymmetry \cite{Steiner80,Hufner,supp} is suppressed compared with a simple metal [Fig. 3(b)].
These features are inconsistent with the usual poorly-screened peaks for the Mn compounds \cite{Biesinger11,Mitra03,MnB24} and instead resemble non‑locally screened (or well-screened) components seen at lower binding energy in doped Mn oxides \cite{Horiba04,Taguchi05,Takubo07,Hufner}.
Actually, cluster calculations that omit Slater exchange terms between Mn 2$p$ and 3$d$ (i.e., $G_{2p3d}^{1,3}$=0) almost reproduce the sharp spectral shapes [bottom of Fig. 3(b)], indicating strong nonlocal screening of the Mn core hole by surrounding conductive bands below $E_\mathrm{F}$ (notably Yb 4$f$/5$d$ and B 2$p$ states) [Fig. 3(c)] \cite{supp}.
Another way to reproduce the spectra is that all of the Slater integrals in the final states are reduced below 10 \% of the atomic values \cite{supp}.
In either case, the absence of multiplet and satellites in HAXPES reflects strong final‑state screening owing to strong hybridization with B 2$p$, and heavier Yb 4$f$ and 5$d$ bands below $E_\mathrm{F}$ rather than a weakly correlated Mn ground state.
These characteristics of the Mn electrons certainly corresponds well to the nonmagnetic nature in the magnetic measurements due to the Kondo effect as like Al-Mn, where the spins in the Mn sites are non-locally-screened and suppressed owing to the hybridization with conductive electrons in the surrounding sites \cite{Steiner80,YoshidaTheory}.

\begin{figure}
	\includegraphics[width=1\linewidth]{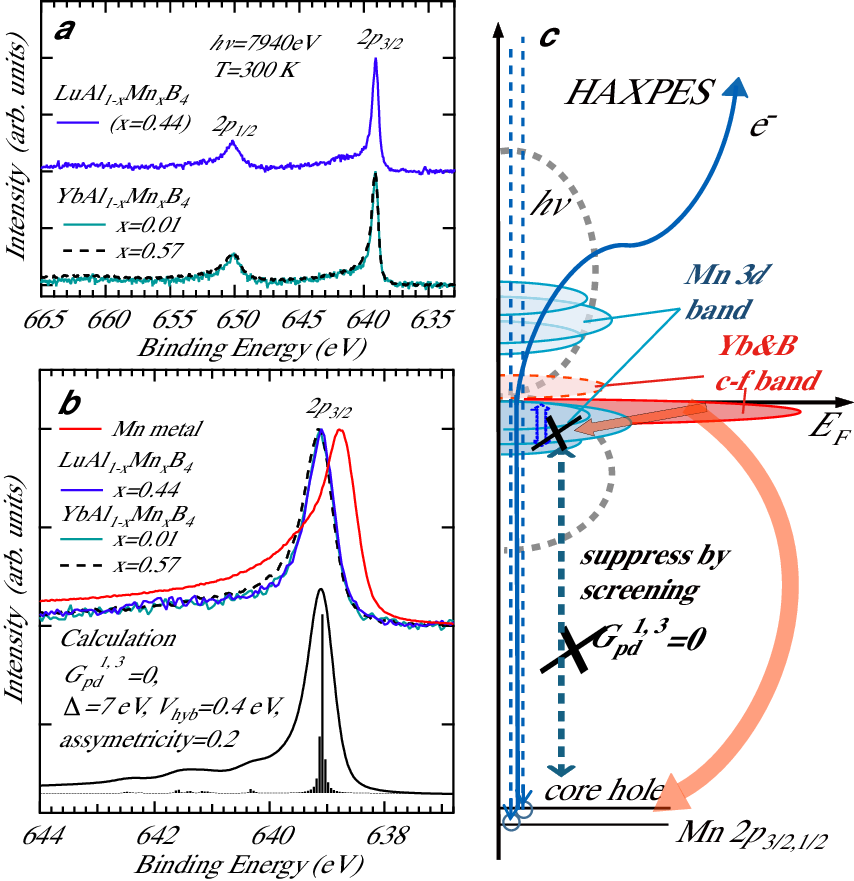}%
	\caption{Mn 2$p$ HAXPES of {$\alpha$-(Yb,Lu)Al$_{1-x}$Mn$_{x}$B$_4$}.
		(a) Spectra of Mn 2$p_{3/2}$ and 2$p_{1/2}$.
		(b) Enlarged view of the Mn 2$p_{3/2}$ peak compared to pure Mn metal. Mn metal spectrum was borrowed from NIMS database \cite{pureMnHAXPES2}.
		The bottom is a calculation for the Mn$^{2+}$ ground state by assuming that the Slater exchange terms $G_{2p3d}^{1,3}$ in the final state are suppressed \cite{supp}.
		(c) Schematic of HAXPES processes showing the excitation from the Mn 2$p$ core-levels.
		The blue and red parabolas indicate the density of states for the Mn$^{2+}$ 3$d$ and Yb$^{2+}$\&B $c$-$f$ bands, respectively.
		The light red blocked arrows indicate the screening effects expected in the HAXPES final state. 
	}
\end{figure}

The contrasting signatures between XAS and HAXPES reported here — multiplet structure in XAS versus strong screening in HAXPES — have not previously been documented in any heavy‑fermion systems.
Both probes, however, reveal complementary aspects of a Kondo‑like correlation at the Mn sites.
The bulk sensitivity of HAXPES with $hv\sim$8 keV and XAS in the TEY mode is comparable ($\sim$a few 10 nm).
The XAS results obtained in the TEY and bulk-sensitive FY modes are very consistent, indicating that this apparent discrepancy between XAS and HAXPES is not due to the surface effect.
In our interpretation, the occupied Mn 3$d$ states lie at or very near $E_\mathrm{F}$ and strongly overlap with heavy Yb and B bands.
Since the Yb (or Lu) states are largely occupied (close to $f^{13}$ or $f^{14}$), the particle–hole symmetry of the band structure is intrinsically broken.
This asymmetry leads to different correlation and screening behavior for occupied versus unoccupied Mn 3$d$ states, explaining why multiplet features appear in XAS (probing unoccupied states), while HAXPES (probing occupied states) is dominated by nonlocal screening.

As illustrated in Fig. 2(b), XAS final states reflect CI and CF effects on the unoccupied Mn$^{2+}$ 3$d$ levels, which lie well above $E_\mathrm{F}$ and do not strongly hybridize with the Yb $c$-$f$ bands. By contrast, the occupied Mn 3$d$ states are located at or very near $E_\mathrm{F}$ and overlap with Yb 4$f$ bands [Fig.3 (c)], which can be confirmed by valence‑band HAXPES (Fig. 4).
Photoemission final states in HAXPES are strongly nonlocally screened because of the large density of states from Yb/Lu 4$f$ and 5$d$ bands below $E_\mathrm{F}$, which hybridize with occupied Mn 3$d$ and B 2$p$ electrons. This strong nonlocal screening reduces the core‑hole effect and suppresses multiplet and charge‑transfer satellites in Mn 2$p$ HAXPES. Although part of the discrepancy between XAS and magnetic probes can be attributed to different timescales, the contrasting final‑state correlation and screening in XAS versus HAXPES has not been fully established. We therefore attribute the pronounced asymmetry between XAS and HAXPES to the intrinsic band structure of $\alpha$-(Yb,Lu)Al$_{1-x}$Mn$_x$B$_4$: because Yb/Lu 4$f$/5$d$ states are nearly filled (close to $f^{13}$ or $f^{14}$), particle–hole symmetry is broken, producing asymmetric correlation and screening for occupied versus unoccupied Mn 3$d$ states.
Since such difference in correlation and screening effects between occupied and unoccupied electrons within the same site has not been suggested, further theoretical and experimental studies will be highly desirable utilized by diverse of spectroscopies including soft x-ray emission, soft XPS, and resonant inelastic scattering as well as XAS and HAXPES for various HF systems.

Finally, we discuss how TM substitution affects the physical properties of $\alpha$-(Yb,Lu)Al$_{1-x}$Mn$_x$B$_4$.
HAXPES indicates that Al remains close to a 3+ valence as shown in SM \cite{supp,NISTXPS,Ma17}, implying that Mn substitutes on Al sites as a nominal Mn$^{2+}$ and introduces hole doping while carrying significant electronic correlations.
Then, this direct hole with hidden electronic and magnetic correlation doping will be main origins of the two-stage valence crossover of the Yb sites (Fig. 1).
Although Mn sites appear nonmagnetic on the muon‑spin timescale, the Mn 3$d$ electrons near $E_F$ strongly hybridize with Yb 4$f$/5$d$ and B 2$p$ states. This direct correlation between Mn and Yb electrons, evidenced by HAXPES, can suppress Yb valence fluctuations and stabilize antiferromagnetic order, accounting for the large $T_N$ ($\sim$20 K) observed for $x\geq$ 0.11. Thus, both the change in carrier concentration (hole doping) and the enhanced hybridization strength likely drive magnetic ordering near the quantum‑critical region, analogous to behavior reported in certain Kondo‑semiconductor systems such as Ce$Tr_2$Al$_{10}$ \cite{Kimura11,CeTr2Al10_1,Ishiga14}.
In these system, CeFe$_2$Al$_{10}$ does not show any magnetic order, CeRu$_2$Al$_{10}$ and CeOs$_2$Al$_{10}$ show high-temperature antiferromagnetic transitions related to charge-density wave instability near QCP.
The hybridization strength has reported to increase in the order of CeRu$_2$Al$_{10}$ $<$ CeOs$_2$Al$_{10}$ $<$ CeFe$_2$Al$_{10}$ \cite{Ishiga14} and the transition temperature is highest in CeOs$_2$Al$_{10}$, which might be induced by substitution of hidden but strong correlation from the TM sites in similar to the present system. 

\begin{figure}
	\includegraphics[width=1\linewidth]{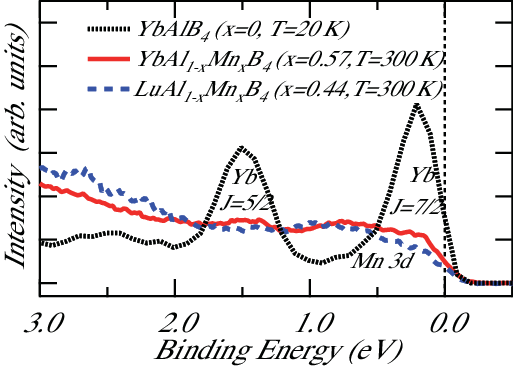}%
	\caption{
		Valence band HAXPES of {$\alpha$-(Yb,Lu)Al$_{1-x}$Mn$_{x}$B$_4$}.
		The spectrum of {$\alpha$-YbAlB$_4$} ($x$ = 0) is replicated from Ref. \onlinecite{Okawa10}. 
	}
\end{figure}

In summary, XAS and HAXPES demonstrate that Mn in heavy fermion $\alpha$-(Yb,Lu)Al$_{1-x}$Mn$_{x}$B$_4$
 is best described as a correlated, high-spin Mn$^{2+}$ that hybridizes strongly with B 2$p$ orbitals and, via further coupling to Yb 4$f$/5$d$ states with each other, acquires heavy-electron character near $E_F$.
Mn 2$p$ XAS and HAXPES reveal complementary but contrasting signatures of Kondo-like correlations, which we attribute to particle-hole asymmetry in the correlated electronic structure and to different final-state screening in the two spectroscopies. Substitution of Al$^{3+}$ by correlated Mn$^{2+}$ effectively hole-dopes the system and provides a route to tune the electronic structure and Yb valence across the observed two-stage large crossover.

Reliable correlations between the valence electrons in metal site and surrounding itinerant ones have been argued to be the origin of various HF quantum behaviors from the early researches but never been clearly established in many systems in despite being based on the diverse experimental studies.
While {$\alpha$-(Yb,Lu)Al$_{1-x}$Mn$_{x}$B$_4$} exhibits an apparent broken particle-hole symmetry, such hidden but strongly correlated TM sites might exist in large number of itinerant electron systems.
The present results will re-stimulate experimental and theoretical researches for the screening and correlation effects in each final state on various spectroscopies, as well as the material control and development for a wide range of itinerant systems utilized by the direct carrier and correlation doping via the substitution of TM sites with hidden but strong correlations.

\section*{Acknowledgements}

XAS measurements at the Mn $L_{2,3}$ edge were performed under the approval of the Photon Factory Program Advisory Committee (Proposals No. 2015G556) at the Institute of Material Structure Science, KEK, and Synchrotron Radiation Research Organization, the University of Tokyo (No. 2014B7401, 2015A7401). 
HAXPES measurements were performed under the approval of SPring-8 (Proposals No. 2015B1699). 
XAS measurements at the Mn $K$ and Yb $L_3$ edges were performed under the approval of SPring-8 (Proposals No. 2014A1664).
This work was supported by the Japan Society for the Promotion of Science (JSPS) of JST-MIRAI Program (JPMJMI20A1), JST-ASPIRE Program (JPMJAP2317) Kakenhi (Grant No. JP25H01250).

\bibliography{SIYbXAS2.bib}

\clearpage


\section*{Supplementary material}

# Supplementary Material

# Re-examination of electronic structure of dilute Kondo transition-metal ions substituted into a Heavy Fermion compound

Kou Takubo,[1] Shintaro Suzuki,[2] Kohei Yamamoto,[2,3] Kohei Yamagami,[2,4] Masafumi Horio,[2] Toshiaki Ina,[4] Kiyohumi Nitta,[4] Masaichiro Mizumaki,[5] Eiji Ikenaga,[4] Yosuke Matsumoto,[2] Hiroki Wadati,[2,6,7] and Satoru Nakatsuji[8,2,9,10]

[1]Department of Applied Physics, Waseda University, Tokyo 169-8555, Japan
[2]Institute for Solid State Physics (ISSP), University of Tokyo, Kashiwa, Chiba 277-8581, Japan
[3]NanoTerasu Center, National Institutes for Quantum Science and Technology, Sendai, Miyagi 980-8572, Japan
[4]JASRI, SPring-8, Kouto, Mikazuki, Sayo, Hyogo 679-5198, Japan
[5]Faculty of Science, Kumamoto University, 2-39-1 Kurokami Chuou-ku, Kumamoto 860-8555, Japan
[6]Department of Material Science, Graduate School of Science, University of Hyogo, Ako, Hyogo 678-1297, Japan
[7]Institute of Laser Engineering, Osaka University, Suita, Osaka 565-0871, Japan
[8]Department of Physics, University of Tokyo, Bunkyo-ku, Tokyo 113-0033, Japan
[9]Trans-scale Quantum Science Institute, University of Tokyo 113-0033, Bunkyo-ku, Tokyo, Japan
[10]Institute for Quantum Matter and Department of Physics and Astronomy, Johns Hopkins University, Baltimore, MD 21218, USA

## 1. XAS at Yb $L_3$ and Mn $K$ edges

Previous studies on $\alpha$-YbAl$_{1-x}$Fe$_x$B$_4$ suggested that a quantum valence transition underlies the crossover from a valence-fluctuating metal to an antiferromagnetic semiconducting phase induced by transition-metal substitution [1]. A sharp valence crossover was observed for $\alpha$-YbAl$_{1-x}$Fe$_x$B$_4$ at quantum critical point (QCP) $x_c \sim 0.014$, where the valence exhibits a sudden increase by 0.03, corresponding to a 3% increase in the $4f$ hole density. The system seems to be controlled from a 'mixed-valent regime' to a 'Kondo regime' through a valence crossover surface by the tuning of chemical pressure or cell volume induced by the Fe or Mn substitution $x$ [1, 2].

To clarify the relationship between the phase diagram and the Yb valence in $\alpha$-YbAl$_{1-x}$Mn$_x$B$_4$, we measured Yb $L_3$-edge x-ray absorption spectra on powdered bulk samples in transmission geometry. The Yb $L_3$ spectra were fitted phenomenologically with two Lorentzian components representing Yb$^{2+}$ and Yb$^{3+}$, following established procedures [3–5]; the spectra were normalized over 8920–8970 eV and the Yb valence was estimated from the integrated area ratio of these components. The resulting valences as a function of Mn concentration $x$ and temperature $T$ are presented in Fig. 1 (main text) and Fig. S1(c), respectively.

The valences for $x = 0.0$ and 0.04 increase from $\sim$2.73+ to $\sim$2.83+ with increasing temperature from $T = 8$ to 300 K. Namely, a valence fluctuation is associated with $x = 0$ and 0.04, and these results are qualitatively consistent with those of previous studies for $x = 0.0$ [1, 5]. The peculiar kinks are observed in the temperature dependence of approximately 35 K for all samples. These may correspond to the scale of the Kondo temperature evaluated for $x > 0.11$ ($T_K \sim 40$ K), although $T_K$ is around 100 K for $x = 0.0$ [2]. In contrast, the Yb$^{2+}$ states seem to be suppressed in higher doped samples ($x \geq 0.11$) [See Fig. S1(b)]. The valence exhibits a further increase by $\sim$0.11 for $x \sim 0.07$ (between 0.04 and 0.11) at $T = 8$ K, as shown in Fig. 1 and S1(c). The valences for $x \geq 0.11$ reach $\sim$2.9+, close to 3+, and their fluctuations are mostly suppressed in the Kondo semiconducting phase. This steep valence crossover may be qualitatively similar to that observed for $\alpha$-YbAl$_{1-x}$Fe$_x$B$_4$ at around $x_c \sim 0.014$, associating with first-order valence transition (FOVT) [1]. However, the total increase by $\sim$0.15 across the two-stage crossovers ($x = \sim 0.009$ and 0.07) for $\alpha$-YbAl$_{1-x}$Mn$_x$B$_4$ is much larger than that by $\sim$0.03 at FOVT for $\alpha$-YbAl$_{1-x}$Fe$_x$B$_4$. This corresponds to a 15% increase of $4f$ hole density and amounts other FOVTs found in the prototypical systems [4]. Therefore, a drastic quantity instead of the chemical pressure or cell volume proposed for $\alpha$-YbAl$_{1-x}$Fe$_x$B$_4$ is expected for the Mn substitution in $\alpha$-YbAl$_{1-x}$Mn$_x$B$_4$ as discussed in the main text, just relating to the origin of the highest $T_N \sim 20$ K for $x \geq 0.11$ (Fig. 1).

For the Mn sites, bulk-sensitive XAS at the Mn $K$ edge was also performed in transmission geometry, in addition to XAS at the Mn $L_{2,3}$-edges shown in the main article. The Mn $K$-edge peak energy in $\alpha$-YbAl$_{1-x}$Mn$_x$B$_4$ is shifted to higher energy relative to Mn foil and is comparable to or slightly higher than MnO (Mn$^{2+}$) and Mn$_2$O$_3$ (Mn$^{3+}$) as shown in Fig. S1(d,e). Although the leading edge near $\sim$6540 eV is almost similar to metallic Mn, the spectral weight in the 6544–6560 eV region is much suppressed compared to MnO and Mn foil. These features indicate that Mn in this compound is best described as nominally 2+ or higher while retaining metallic character in the $K$-edge XAS.

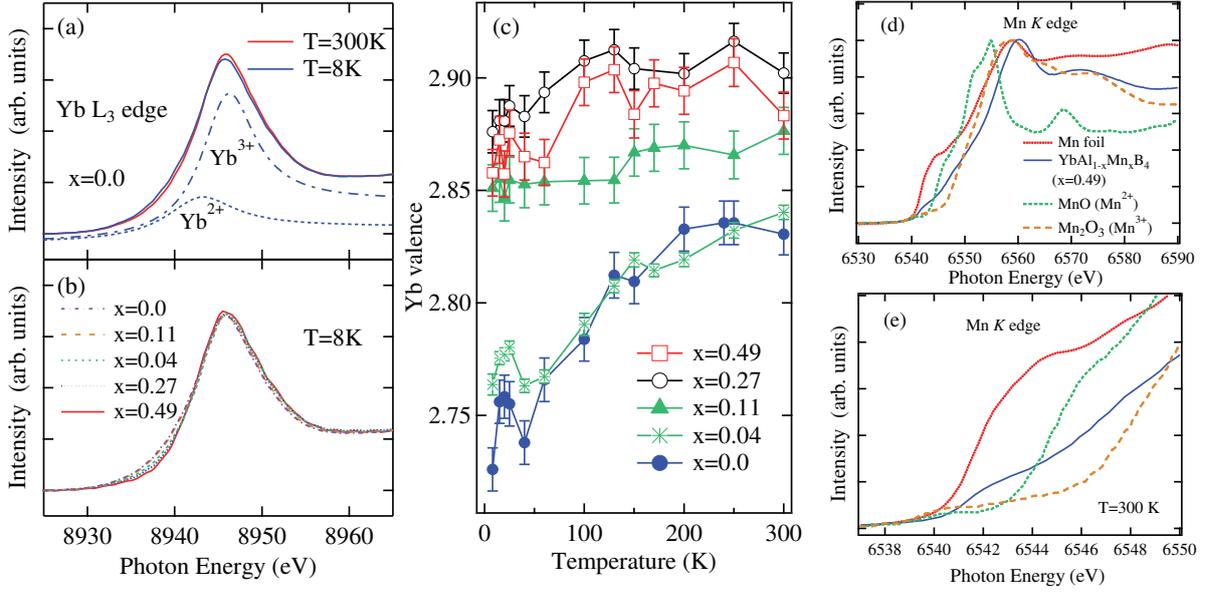

FIG. S1. XAS at the Yb $L_3$ and Mn $K$ absorption edges of $\alpha$-YbAl$_{1-x}$Mn$_x$B$_4$. (a) Spectra at the Yb $L_3$ edge for $x = 0.0$ at $T$ = 8 and 300 K. The fitting curves for the Yb$^{2+}$ and Yb$^{3+}$ components at $T$=8 K are indicated by dotted lines. (b) Spectra at the Yb $L_3$ for various Mn concentrations at $T$=8 K. (c) Temperature dependence of the valence at the Yb sites. (d) Spectra at the Mn $K$-edge of $\alpha$-YbAl$_{1-x}$Mn$_x$B$_4$ ($x$=0.49) and a Mn foil. The spectra of MnO and Mn$_2$O$_3$ are replicated from Ref. [6]. (e) Spectra at the Mn $K$-edge, zoomed into the low energy region.

## 2. Charge-transfer multiplet calculations of XAS and HAXPES

The XAS spectra at the Mn $L_{2,3}$ ($2p_{1/2,3/2} \to 3d$) absorption edges of $\alpha$-(Yb,Lu)Al$_{1-x}$Mn$_x$B$_4$ were simulated by a charge-transfer multiplet calculation using Quanty [7]. The codes were made by referring from the input files for $D_{3h}$ and $D_{4h}$ symmetries of Crispy [8]. The local crystal-field (CF) around the Mn site was modeled by an Mn-B$_{10}$ pentagonal prism having $D_{5h}$ symmetry [Inset of Fig. S2(a)], and the energy splittings in $D_{5h}$ were described by the $B_2^0$ and $B_4^0$ Wybourne coefficients for the spherical harmonics.

Figure S2(a) shows the experimental and calculated spectra for the Mn$^{1+}$ ($3d^6$), Mn$^{2+}$ ($3d^5$), and Mn$^{3+}$ ($3d^4$) ground states. The Coulomb interactions of $U_{dd}$ and $U_{pd}$ were set to 6.0 and 7.2 eV, respectively. The Slater integrals $F^2$, $F^4$, $G^1$, and $G^3$ were reduced to 85 % of the atomic values [9]. Best fits were obtained for the Mn$^{2+}$ ground state with Wybourne coefficients of $B_2^0 = -1.0$ eV and $B_4^0 = 0.2$ eV, Lorentzian widths of 0.4 eV ($L_3$) and 0.8 eV ($L_2$), and a small Gaussian broadening (0.05 eV). The charge transfer and hybridization energies ($\Delta$ and $V_{hyb}$) from the Mn $3d$ to surrounding B $2p$ orbitals were neglected in the calculation of Figs. S2(a) to (d), although these effects are found to be not significant for the present XAS in the region of $\Delta \geq 4$ eV and $V_{hyb} \leq 0.5$ eV, consistent with a high-spin Mn$^{2+}$ ground state, as indicated later in Fig. S2(e).

Figure S2(b) gives calculated XAS spectra of the Mn$^{2+}$ state with the various reduction factor (RF) of the Slater integrals. While no CF splitting nor charge-transfer between Mn $3d$ and ligand B $2p$ orbitals is included in the calculation of Fig. S2(b) ($B_2^0$, $B_4^0$, and $V_{hyb}$= 0), the multiplet structures in the $L_3$ region are essentially reproduced by assuming the ground state of the high-spin Mn$^{2+}$ with RF around 85 %. On the other hand, if the effects of the CF splitting were not properly considered, the detailed shape of the experimental spectrum was difficult to be reproduced in particular for the $L_2$-edge region where the gap between two distant peaks was mostly filled. Figures S2(c) and S(d) give $B_2^0$ and $B_4^0$ dependencies of the spectra, respectively. Only when $B_2^0$ is around $-1.0$ eV with considerable positive value of $B_4^0$, the overall spectral shape including the $L_2$ structure is reproduced where the gap between the two peaks is filled. In addition, figure S2(e) shows the dependencies of the spectra on the charge transfer $\Delta$ and hybridization energy $V_{hyb}$ between the Mn $3d$ and B $2p$ states. Although these effects are not evident in the present calculation at $\Delta \geq 4.0$ eV, the overall XAS results suggest that the ligand CF and/or correlation between the surrounding sites is appeared around the high-spin Mn$^{2+}$ states ($3d^5$) in $\alpha$-(Yb,Lu)Al$_{1-x}$Mn$_x$B$_4$.

In this CF parameter, the energies of the Mn $3d$ orbitals are split in the order of $A_1' \leq E_1' < E_2'$ states, while the out-of-plane $A_1'$ and $E_1'$ states seems to be almost degenerate [Inset of Fig. S2(a)]. The one-electron orbital energies of $A_1'$, $E_1'$, and $E_2'$ in the $D_{5h}$ symmetry can be expressed as the $2/7B_2^0 + 6/35B_4^0$, $1/7B_2^0 - 4/35B_4^0$, and

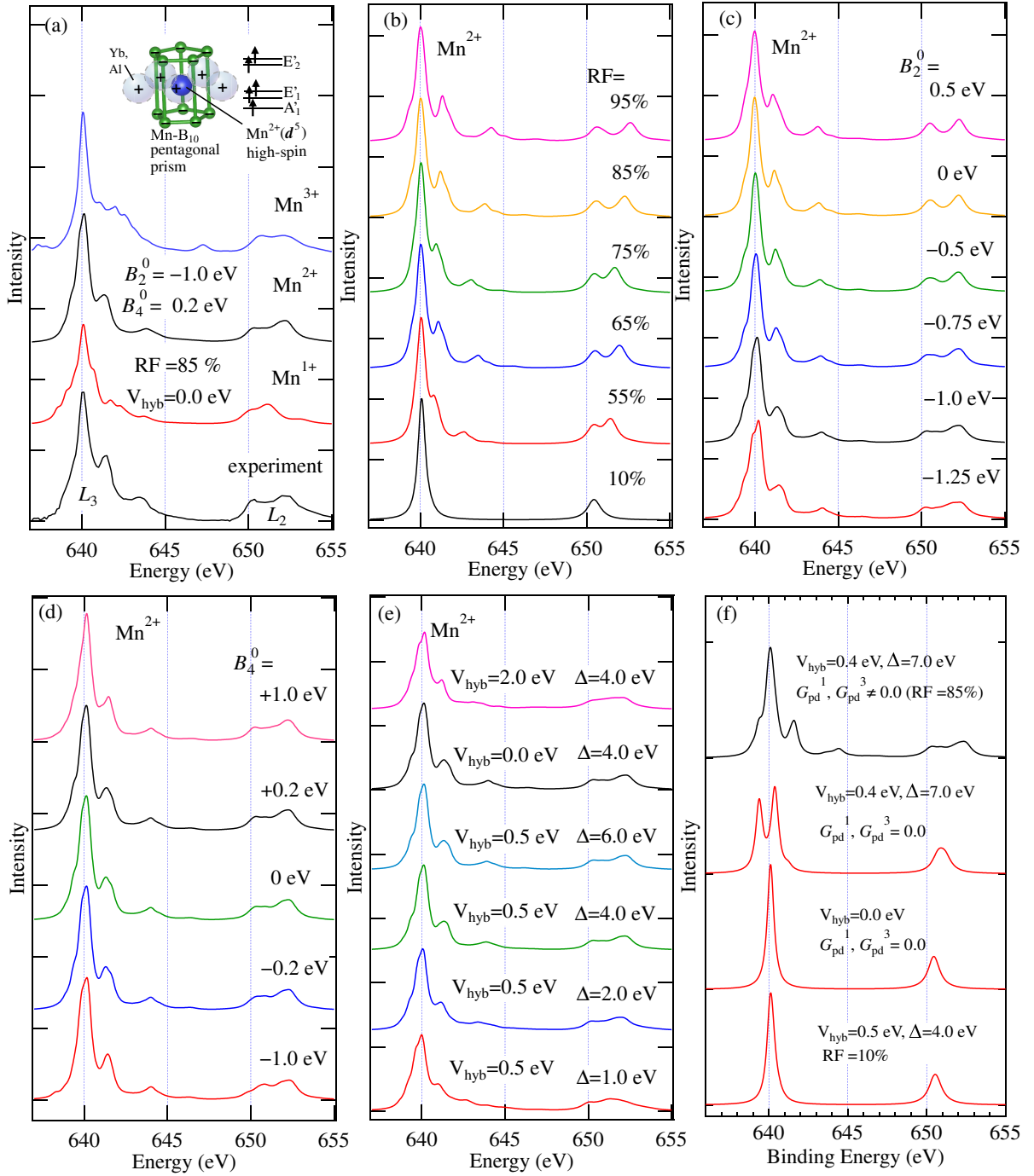

FIG. S2. Calculation of the Mn $L_{2,3}$-edge XAS. (a) Calculated XAS spectra for $Mn^{1+}$ ($d^6$), $Mn^{2+}$ ($d^5$), and $Mn^{3+}$ ($d^4$) ground states. The bottom is the experimental data of $\alpha$-YbAl$_{1-x}$Mn$_x$B$_4$ ($x$=0.27). Inset shows the pentagonal prism of Mn-B$_{10}$ cluster and expected energy-splitting. The best-fit spectrum for the $Mn^{2+}$ ground state is also shown in Fig. 2 of the main text. (b) Calculated XAS spectra for $Mn^{2+}$ state at various RF. Here, the CF splitting and charge transfer between Mn $3d$ and B $2p$ orbitals were neglected. (c) $B_2^0$ dependence of the spectra for $Mn^{2+}$ at RF=85 %. Ratio of $B_4^0/B_2^0$ was fixed to $-0.2$. (d) $B_4^0$ dependence of the spectra for $Mn^{2+}$ at RF=85% and $B_2^0=-1.0$ eV. (e) Effects of the hybridization and charge-transfer between the Mn $3d$ and B $2p$ orbitals. (e) Effects of final-state effects in which the Slater exchange terms between Mn core-hole $2p$ and $3d$ or all of the Slater integrals are screened ($G_{pd}^1$ and $G_{pd}^3 = 0$ or RF$_{\text{final}} = 10$ %).

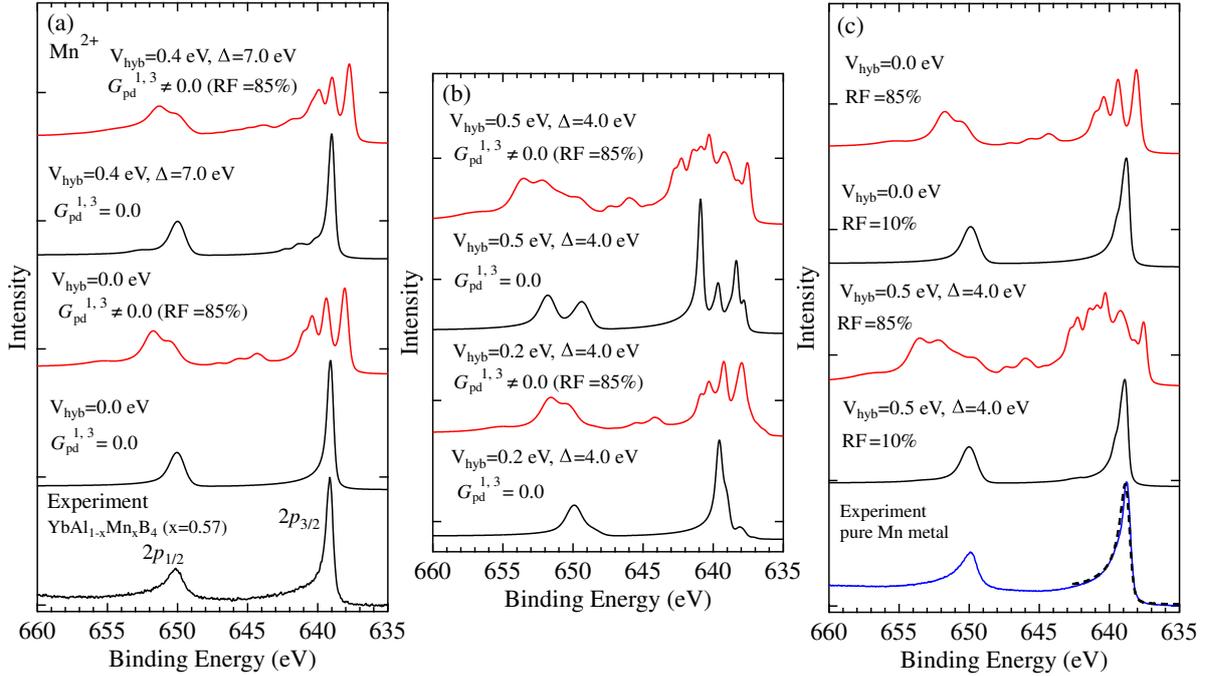

FIG. S3. Calculation of Mn 2p HAXPES of the $Mn^{2+}$ ground state. $B_0^2$ and $B_0^4$ are $-1.0$ and $0.2$ eV, respectively. (a,b) Effects of core-hole Slater exchange terms of $G_{pd}^1$ and $G_{pd}^3$ for the calculation with (a) rather moderate and (b) strong hybridization between Mn 3d and B 2p. The bottom in (a) is the experimental data of $\alpha$-YbAl$_{1-x}$Mn$_x$B$_4$ ($x$=0.57). (c) Effects of RF$_{\text{final}}$ for the calculation. The bottom is the experimental data of pure Mn metal [11]. Here, the widths of Lorentzian and Gaussian, and asymmetric factor were settled to 0.2 eV, 0.1 eV, and 0.2, respectively for the calculation of $\alpha$-YbAl$_{1-x}$Mn$_x$B$_4$. On the other hand, the asymmetric factor for the pure metallic Mn spectrum [11] was estimated around 0.45 .

$-2/7 B_2^0 + 1/35 B_4^0$, respectively. In $\alpha$-YbAl$_{1-x}$Mn$_x$B$_4$, the four and one sides of the Mn-B$_{10}$ prism face the Yb and Al cations, respectively, in which the Mn and huge Yb (and Al) cations are in similar $ab$ plane only about 2.9 Å apart [See inset of Fig. 1] [2, 10]. On the other hand, the length of $c$-axis is ~3.5 Å. Namely, the pentagonal prism of Mn-B$_{10}$ is elongated along the $c$-axis. Therefore, it is reasonable that the energy of the in-plane orbitals $E_2'$ is higher than those of out-of-plane orbitals $A_1'$ and $E_1'$, resulting the negative value of $B_2^0$. On the other hand, the anion B sites in the upper and lower layers are in a direction close to the $yz$- and $zx$-type orbitals of the $E_2'$ state rather than the $3z^2$-type orbital of the $A_1'$ state, then $B_4^0$ will become positive which is consistent with the XAS results. In addition, the short distance of ~2.9 Å between Mn and Yb (or Al) cations can cause direct correlation between the electrons in these sites. While the present calculation for the Mn-B$_{10}$ cluster does not incorporate this correlation directly, this tendency in XAS can be consistently expressed by CF with these $B_2^0$ and $B_4^0$ values.

In stark contrast, conventional cluster models including core-hole exchange (Slater exchange terms $G_{pd}^{1,3}$) failed to reproduce the experimentally sharp, symmetric peaks for Mn 2p hard x-ray photoemission spectra (HAXPES) (Shown in reds in Fig. S3). The spectra can be well reproduced either by (i) omitting the Slater exchange between Mn 2p and 3d in the final state ($G_{pd}^{1,3}$=0) with rather moderate hybridization of $\Delta \geq 7$ eV and $V_{hyb} \leq 0.4$ eV [Shown in blacks in Fig. S3(a)], or (ii) strongly reducing all final-state Slater integrals (to $\leq 10$ % of atomic values) [Shown in blacks in Fig. S3(c)]. The latter situation would be inconsistent with the XAS results for the correlated unoccupied $Mn^{2+}$ 3d states [Inset of Fig. S2(a)], while HAXPES might possibly reflect electronic correlations of the final states in the occupied region. In either case, both scenarios imply strong nonlocal screening of the Mn core hole by surrounding conduction bands below $E_F$ (notably Yb 4f/5d and B 2p) [12, 13], consistent with the HAXPES observation of well-screened components and the absence of pronounced multiplet satellites as discussed in main text. The present HAXPES results will suggest a hidden but rather direct correlation between the Mn 3d and heavier Yb 4f/5d states in the occupied region beyond the cluster of Mn-B$_{10}$ due to the very short distance of ~2.9 Å between these sites, just corresponding to the screened, nonmagnetic nature of the Mn electrons by the Kondo effect. It should be noted that these final-state screening was apparently not observed on soft x-ray absorption at the Mn $L$-edge probing the unoccupied 3d states in contrast, as confirmed in Fig. S2(f).

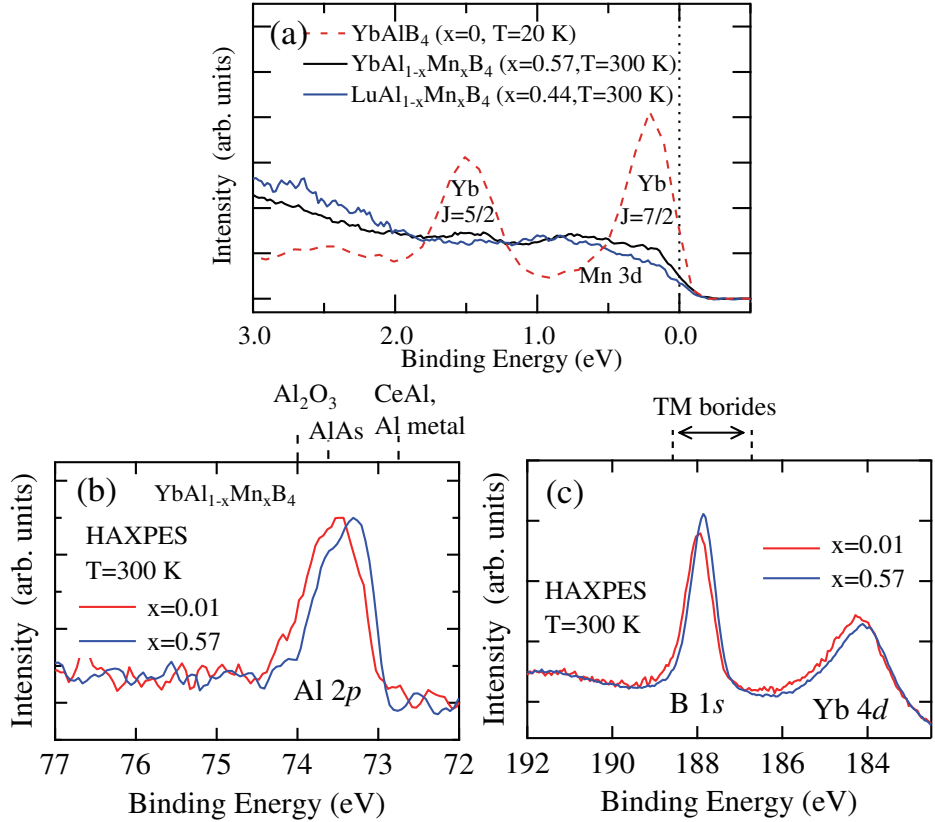

FIG. S4. HAXPES of (a) valence band, (b) Al 2$p$, and (c) B 1$s$ and Yb 4$d$ for $\alpha$-YbAl$_{1-x}$Mn$_x$B$_4$ and $\alpha$-LuAl$_{1-x}$Mn$_x$B$_4$. The spectrum of $\alpha$-YbAlB$_4$ ($x=0$) in (a) is replicated from Ref. 14. The dashed lines above the figures (b) and (c) indicate the binding energies of typical Al and B compounds.

## 3. HAXPES of valence band, and core-levels of Al and B

The correlation between the Yb 4$f$ and 5$d$, and Mn 3$d$ states in the occupied region proposed from the core-level spectroscopy is directly confirmed on the valence-band HAXPES shown in Fig. S4(a) and Fig. 4 of main text. The Mn 3$d$–derived states lie immediately below $E_\mathrm{F}$. In $\alpha$-LuAl$_{1-x}$Mn$_x$B$_4$ (Lu$^{3+}$), the Mn 3$d$ band appears near $E_\mathrm{F}$ without significant Lu 4$f$ contribution, whereas in Yb-containing samples sharp Yb 4$f$ features overlap the Mn 3$d$ states, indicating the strong Mn–Yb hybridization.

Figures S4(b) and S4(c) show the Al 2$p$ and B 1$s$ HAXPES spectra, respectively. The Al 2$p$ core levels are centered near 73.3–73.5 eV, consistent with an Al$^{3+}$-like chemical environment (comparable to AlAs or Al$_2$O$_3$) [15], while B 1$s$ binding energies of 187.8–187.9 eV fall within the range typical for transition-metal borides [16]. A small negative shift of $-0.1$ eV in Al 2$p$ and B 1$s$ between $x=0.01$ and 0.57 is observed, consistent with hole doping introduced by the Mn substitution. By contrast, the Mn 2$p$ HAXPES shows negligible energy shift ($<0.05$ eV) across this composition range (Fig. 3 in main text), just reflecting robust final-state screening.


\clearpage


\end{document}